\documentclass[numbers,preprint,3p]{elsarticle}
\pdfoutput=1 

\usepackage{amssymb}

\usepackage{amsmath}

\usepackage{nomencl}
\makenomenclature
\usepackage[colorlinks=true,linkcolor=black, citecolor=blue, urlcolor=blue]{hyperref}

\usepackage{tabularx}
\usepackage{multirow}
\usepackage{makecell}

\DeclareMathOperator*{\argminB}{argmin}

\usepackage[final]{changes}
\usepackage{floatrow}
\usepackage{lineno}
\newcommand*\patchAmsMathEnvironmentForLineno[1]{%
  \expandafter\let\csname old#1\expandafter\endcsname\csname #1\endcsname
  \expandafter\let\csname oldend#1\expandafter\endcsname\csname end#1\endcsname
  \renewenvironment{#1}%
     {\linenomath\csname old#1\endcsname}%
     {\csname oldend#1\endcsname\endlinenomath}}%
\newcommand*\patchBothAmsMathEnvironmentsForLineno[1]{%
  \patchAmsMathEnvironmentForLineno{#1}%
  \patchAmsMathEnvironmentForLineno{#1*}}%
\AtBeginDocument{%
\patchBothAmsMathEnvironmentsForLineno{equation}%
\patchBothAmsMathEnvironmentsForLineno{align}%
\patchBothAmsMathEnvironmentsForLineno{flalign}%
\patchBothAmsMathEnvironmentsForLineno{alignat}%
\patchBothAmsMathEnvironmentsForLineno{gather}%
\patchBothAmsMathEnvironmentsForLineno{multline}%
}

\begin{document}

\begin{frontmatter}

\title{Predicting wave heights for marine design by prioritizing extreme events in a global model}

\author[label1,label2]{Andreas F. Haselsteiner\corref{cor1}}
\address[label1]{University of Bremen, BIK -- 	
Institute for Integrated Product Development, 28359 Bremen, Germany}
\address[label2]{ForWind -- Center for Wind Energy Research of the Universities of Oldenburg, Hannover and Bremen}
\cortext[cor1]{Corresponding author}
\ead{a.haselsteiner@uni-bremen.de}

\author[label1,label2]{Klaus-Dieter Thoben}
\ead{thoben@uni-bremen.de}

\begin{abstract}
In the design process of marine structures like offshore wind turbines the long-term distribution of significant wave height needs to be modelled to estimate loads. This is typically done by fitting a translated Weibull distribution to wave data. However, the translated Weibull distribution often fits well at typical values, but poorly at high wave heights such that extreme loads are underestimated. Here, we analyzed wave datasets from six locations suitable for offshore wind turbines. We found that the exponentiated Weibull distribution provides better overall fit to these wave data than the translated Weibull distribution. However, when the exponentiated Weibull distribution was fitted using maximum likelihood estimation, model fit at the upper tail was sometimes still poor. Thus, to ensure good model fit at the tail, we estimated the distribution's parameters by prioritizing observations of high wave height using weighted least squares estimation. Then, the distribution fitted well at the bulks of the six datasets (mean absolute error in the order of 0.1\,m) and at the tails (mean absolute error in the order of 0.5\,m). \deleted{Our results suggest that fitting the exponentiated Weibull distribution to wave data can improve the accuracy of design load cases for offshore wind turbines.} \added{The proposed method also estimated the wave height's 1-year return value accurately and could be used to calculate design loads for offshore wind turbines.}
\end{abstract}


\end{frontmatter}


\printnomenclature
\nomenclature{CDF}{Cumulative distribution function}
\nomenclature{DLC}{Design load case}
\nomenclature{ICDF}{Inverse cumulative distribution function}
\nomenclature{MLE}{Maximum likelihood estimation}
\nomenclature{PDF}{Probability density function}
\nomenclature{WLS}{Weighted least squares}


\section{Introduction}
To estimate the loads on a marine structure like an offshore wind turbine, the long-term distribution of environmental variables that describe wave characteristics needs to be modelled. Especially important is the variable significant wave height, which describes the intensity of a sea state. The long-term distribution of significant wave height is typically estimated by fitting a parametric probability distribution to measured or simulated wave data. Then, based on this probability distribution different quantiles are derived and used as design conditions for structural integrity calculations of the marine structure of interest. For example, standards for offshore wind turbines \cite{IEC61400-3-1:2019-04, IECTS61400-3-2:2019-04}, require designers to estimate the 1-year and 50-year return value of significant wave height -- extreme values that are exceeded, on average, every 1 and 50 years, respectively. To calculate these values, designers might use a `global model' or an `event model'. Global models are derived using all available data from long series of subsequent observations while event models are derived from selected extremes of the original dataset \cite[][p. 75]{DNVGL-RP-C205:2017}. 
\par 
These two approaches have different strengths and weaknesses. Global models utilize the complete original dataset and consequently make use of all available information. Further, no preprocessing is required and common parametric distributions can be used. However, time series of significant wave height show strong auto-correlation such that the individual datapoints are not independent and identically distributed (IID condition). Additionally, common fitting approaches like maximum likelihood estimation (MLE) and least squares estimation weight every datapoint equally and thus do not take into account that in marine design high values of significant wave height are especially important. 
\par 
Event models are typically fitted using the peak over threshold method or the block maximum method. In both cases, the original time series are preprocessed and invidiual peaks are identified. These peaks fulfill the IID condition to a much higher degree than the raw time series. However, much less information is used when fitting a distribution to these peaks. Further, event models only describe the upper -- or as a synonym right -- tail of the global distribution of significant wave height. In design, sometimes also quantiles within the bulk of the distribution need to be estimated such that a second model that covers lower quantiles is required. Often, this second model is a global model such that, in one design project, two diverging models for high quantiles might exist.
\par 
Here, we focus on global models. In the past, significant wave height has been modelled using various parametric distributions with two to five parameters: The lognormal distribution \cite{Jasper1956, Ochi1980}, the 2-parameter Weibull distribution \cite{Battjes1972, Ochi1980}, 
the translated Weibull distribution (sometimes simply called `3-parameter Weibull distribution') \cite{Nordenstrom1973, Vanem2019:Uncertainty}, the generalized gamma distribution \cite{Ochi1992}, 
a 3-parameter beta distribution of the second kind (and two similar distributions, which showed worse model fit) \cite{Ferreira1999}, Ochi's four-parameter distribution \cite{Ochi1980} and the `Lonowe distribution' \cite{Haver1985, li:2015} (Tab. \ref{tab:common-models-for-hs}). At the moment, probably the most used distribution to model significant wave height is the Weibull distribution. While some others authors use the 2-parameter Weibull distribution (for example \cite{Vanem2015:ClimateChange, Giske2018}), most authors use the translated Weibull distribution (for example \cite{Bitner-Gregersen2015, Vanem2015:ClimateChange, Orimolade2016, vanem:2016, Bitner-Gregersen2018:OpenSea, Vanem2019:Uncertainty, Velarde2019}).
Certifying organizations also recommend to assume that significant wave height follows a translated Weibull distribution unless data indicate otherwise \cite[][p. 76]{DNVGL-RP-C205:2017}.
\par
\begin{figure}
    \centering
    \includegraphics{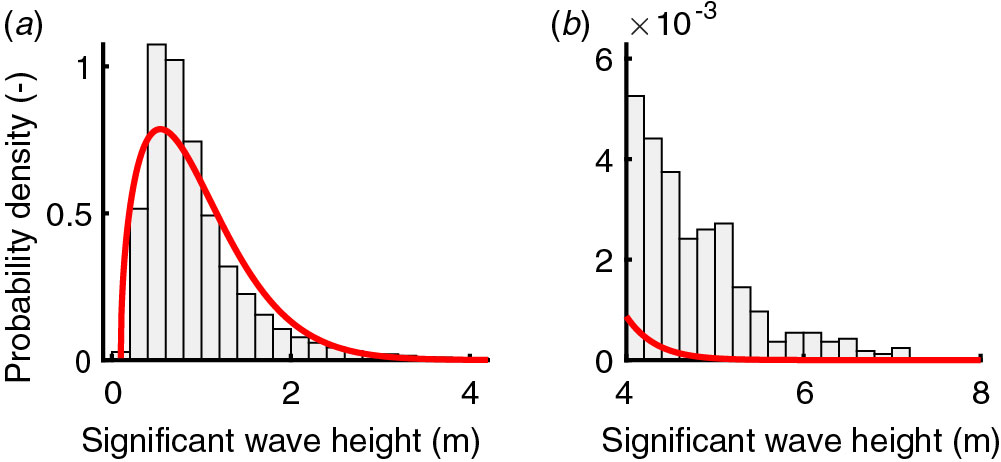}
    \caption{\textbf{Typical model fit of the translated Weibull distribution.} The distribution fits the data relatively well at its body (\textit{a}), but poorly at its tail (\textit{b}).}
    \label{fig:typical-fit-translated-weibull}
\end{figure}

 
However, the translated Weibull distribution often does not fit well at its upper tail and  understimates high quantiles (Fig. \ref{fig:typical-fit-translated-weibull}). Further, some authors have criticized that the distribution's location parameter, which represents a minimum non-zero value, lacks physical meaning since sea states of significant wave height zero exist and represent the calm sea \cite{Ochi1980}. 
\par 
In this paper, we will show that a similar distribution, the exponentiated Weibull distribution, provides better model fit to significant wave height data than the translated Weibull dsitribution. The exponentiated Weibull distribution has three parameters as well and consequently does not increase model complexity. Instead of a location parameter, the distribution has a second shape parameter, which offers the flexibility that is required to ensure good model fit at both, the distribution's bulk and the tail. 

\begin{table*}[]
    \centering
    \begin{tabular}{l l l l}
         Distribution & Nr. of parameters & References that used it for $H_s$ & Proposed in\\
         \hline
         Lognormal & 2 & \cite{Jasper1956, Ochi1980} & 1956 \cite{Jasper1956}\\
         2-parameter Weibull & 2 & \cite{Battjes1972, Ochi1980, Vanem2015:ClimateChange} & 1972 \cite{Battjes1972}\\
         Translated Weibull & 3 & \cite{Nordenstrom1973, Bitner-Gregersen2015, Vanem2015:ClimateChange, Orimolade2016, vanem:2016, Vanem2019:Uncertainty, Velarde2019} & 1973 \cite{Nordenstrom1973}\\
         Generalized gamma & 3 & \cite{Ochi1992} & 1992 \cite{Ochi1992}\\
         3-parameter beta (second kind) & 3 & \cite{Ferreira1999} & 1999 \cite{Ferreira1999}\\
         Ochi distribution & 4 & \cite{Ochi1980} & 1980 \cite{Ochi1980}\\
         Lonowe distribution & 4-5 & \cite{Haver1985,li:2015} & 1985 \cite{Haver1985}\\
    \end{tabular}
    \caption{\textbf{Distributions that have been used to model the long-term distribution of significant wave height.}}
    \label{tab:common-models-for-hs}
\end{table*}

\section{Exponentiated Weibull distribution}
The exponentiated Weibull distribution is a generalization of the common 2-parameter Weibull distribution. It has been proposed by Mudholkar and Srivastava \cite{Mudholkar1993} to model nonmonotone failure rates and has subsequently been used in a variety of contexts (for a review, see Nadarajah \textit{et al.} \cite{Nadarajah2013}). It extends the 2-parameter Weibull distribution with a second shape parameter, $\delta$, that comes as an exponent of the cumulative distribution function (CDF):
\begin{equation}
    F(x) = \left[1 - e^{-(x / \alpha)^\beta}\right]^\delta
\end{equation}
for $x>0$, $\alpha>0$, $\beta>0$ and $\delta>0$. In the case of $\delta=1$ the exponentiated Weibull distribution becomes the 2-parameter Weibull distribution. For comparison, the translated Weibull distribution's CDF, which has a location parameter, $\gamma$, instead of a second shape parameter reads
\begin{equation}
    F(x) = 1 - e^{-[(x - \gamma) / \alpha]^\beta}.
\end{equation}

\section{Material and methods}
To assess whether the exponentiated Weibull distribution represents a  better model for significant wave height, we analyzed hourly time series of wave data of six locations. We considered three models: (i) the translated Weibull distribution with its parameters estimated using maximum likelihood estimation (MLE), (ii) the exponentiated Weibull distribution, fitted using maximum likelihood estimation and (iii) the exponentiated Weibull distribution, fitted using weighted least squares (WLS) estimation. To assess the goodness of fit of the three models, we computed the mean absolute error (MAE) between the models' predictions and the observations. Further, we computed 1-year and 50-year return values and visually inspected quantile-quantile (QQ) plots. In the following, we describe the datasets, the parameter estimation and the goodness of fit assessment in detail.
\subsection{Datasets}
\begin{figure}
    \centering
    \includegraphics[]{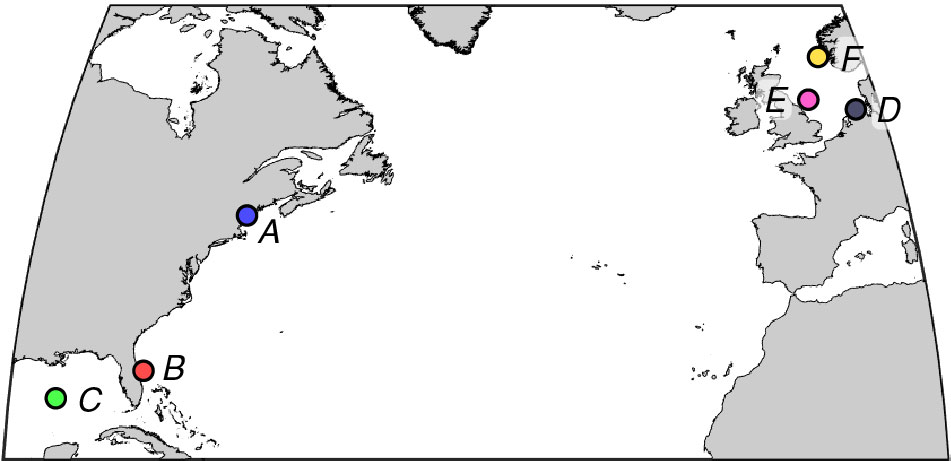}
    \caption{\textbf{Locations of the used datasets.}}
    \label{fig:locations-datasets}
\end{figure}

We used six datasets of significant wave height (Tab. \ref{tab:datasets}). Three datasets ($A$, $B$ \& $C$) were derived from moored buoys off the US East Coast and three datasets were gathered from a hindcast that covers the North Sea ($D$, $E$ \& $F$; Fig. \ref{fig:locations-datasets}). The buoy datasets were recorded by the National Data Buoy Center (NDBC; \cite{NDBC2009}) and were downloaded from \url{www.ndbc.noaa.gov}. They cover the time between January 1st, 1996 and December 31st, 2005. However, the buoys did not measure the complete duration such that these datasets hold between 81,749 (dataset $C$) and 83,917 (dataset $A$) hourly measurements. Datasets $D$, $E$ and $F$ were simulated in the hindcast `coastDat-2' \cite{Groll2016,Groll2017} and cover the complete time between January 1st, 1965 and December 31st, 1989. Additionally, for each location we retained some data (datasets $A_r, B_r, C_r, D_r, E_r$ \& $F_r$) to assess how well the fitted distributions can predict a future time period. The NDBC datasets were preprocessed: we filtered out time periods when no measurements have been conducted, calculated significant wave height from the spectral energy, and created consistent hourly time series by combining 30-minute sea states to hourly sea states when sea states with a duration of 30 minutes instead of 60 minutes were recorded. No preprocessing has been performed on the coastDat-2 datasets. The datasets we used in this paper, are also used in an ongoing benchmarking study on estimating extreme environmental conditions for engineering design \cite{Haselsteiner2019:OMAE}.
\begin{table*}[htb]
    \caption{\textbf{Used datasets of significant wave height.} The buoy data were downloaded from the website of the National Buoy Data Center, \url{www.ndbc.noaa.gov}, and the hindcast samples were derived from the coastDat-2 hindcast \cite{Groll2016}. $n$ = Number of observations.}
    \centering
    \begin{tabularx}{\textwidth}{l l r l l}
        \hline
        \textbf{Dataset}  
        & \textbf{Duration} 
        & \textbf{\textit{n}}
        & \textbf{Site}            
        & \textbf{Data source}  \\      

         \hline
         $A$ & Jan. 1996 to Dec. 2005 
         & 82,805 &
         \multirow{2}{*}{\makecell[l]{\href{https://goo.gl/maps/Dw9TjLwnvZp}{43.525\,N\,70.141\,W} (off Maine coast)}}
         & \multirow{2}{*}{buoy 44007}
         \\
         $A_r$
         & Jan. 2006 to Oct. 2017 
         & 92,515 & & \\
         $B$ & Jan. 1996 to Dec. 2005 
         & 83,917 & \multirow{2}{*}{\makecell[l]{\href{https://goo.gl/maps/ija4Ckko2TC2}{28.508\,N\,80.185\,W} (off Florida coast)}}
         & \multirow{2}{*}{buoy 41009}\\
         $B_r$
         & Jan. 2006 to Jul. 2017 
         & 91,403 & & \\
         $C$ & Feb. 1996 to Dec. 2005  
         & 81,749 
         & \multirow{2}{*}{\makecell[l]{\href{https://goo.gl/maps/PC9PpQmBpDTHaKPdA}{25.897\,N\,89.668\,W} (Gulf of Mexico)}}
         & \multirow{2}{*}{buoy 42001}\\
         $C_r$
         & Jan. 2006 to Jun. 2018 
         & 93,571 & & \\
         $D$
         & Jan. 1965 to Dec. 1989   
         & 219,144
         & \multirow{2}{*}{\makecell[l]{\href{https://goo.gl/maps/rHvi6VzSaBx}{54.000\,N\,6.575\,E} (off German coast)}}
         & \multirow{2}{*}{{hindcast}}\\
         $D_r$
         & Jan. 1990 to Dec. 2014 
         & 219,144 & &\\
         $E$
         & Jan. 1965 to Dec. 1989  
         & 219,144
         & \multirow{2}{*}{\makecell[l]{\href{https://goo.gl/maps/y5W7KxphYK12}{55.000\,N\,1.175\,E}, (off UK coast)}}     
         & \multirow{2}{*}{{hindcast}}\\
         $E_r$
         & Jan. 1990 to Dec. 2014 
         & 219,144 & & \\
         $F$ & Jan. 1965 to Dec. 1989 
         & 219,144
         & \multirow{2}{*}{\makecell[l]{\href{https://goo.gl/maps/osfHEwMAA522}{59.500\,N\,4.325\,E} (off Norwegian coast)}}
         & \multirow{2}{*}{{hindcast}}\\
         $F_r$
         & Jan. 1990 to Dec. 2014 
         & 219,144 & & \\
    \end{tabularx}
    \label{tab:datasets}
\end{table*}

\subsection{Parameter estimation methods}
We estimated the parameters of the translated Weibull distribution using maximum likelihood estimation (MLE). MLE is a standard parameter estimation technique \cite{Scholz2006} and is commonly used in the context of estimating the distribution of significant wave height (see, for example,  \cite{Vanem2019:Uncertainty}). We used Matlab's (Mathworks, USA, version 2019a) function \texttt{MLE.m} to perform the MLE computation. For the second model, an exponentiated Weibull distribution fitted using MLE, we used Matlab's function \texttt{MLE.m} as well. The third model, the exponentiated Weibull distribution whose parameters are fitted using weighted least squares estimation, represents a less typical parameter estimation technique and is therefore explained in detail.
\par 
The goal of this approach is to minimize the sum of the weighted squared deviations between the observed quantiles and the predicted quantiles. Let the set $\{x_i\}_{i=1}^n$ represent the ordered values of a significant wave height sample with $x_1$ representing the lowest measured value and $x_n$ representing the highest measured value, where $n$ represents the length of the sample. Each ordered value, or sample quantile,  $x_i$, has an associated probability $p_i = (i - 0.5) / n$ where $i$ is the index of the ordered value, $i \in [1, n]$. Further, let $\hat{x_i}$ denote the predicted quantile based on an exponentiated Weibull distribution with the parameters $\alpha, \beta, \delta$. Then, the set of parameters that minimizes the sum of the weighted squared deviations between the sample quantiles and the predicted quantiles can be expressed as 
\begin{equation}
    \{\hat{\alpha}, \hat{\beta}, \hat{\delta}\} = \argminB_{\alpha, \beta, \delta} \sum_{i=1}^n w_i(x_i - \hat{x_i})^2,
    \label{eq:argmin-exponentiated}
\end{equation}
\begin{equation}
    \textnormal{where } \hat{x_i} = F^{-1}(\alpha, \beta, \delta; p_i)
\end{equation}
and $F^{-1}$ denotes the inverse cumulative distribution function (ICDF). While in principal many functions for the weights, $w_i$, are possible, here we chose to weight the error based on the squared wave height, 
\begin{equation}
   w_i = \dfrac{x_i^2}{\sum_{i=1}^n x_i^2}.
\end{equation}
Thus, errors between observation and prediction at high wave heights contribute much stronger to the overall error than errors at low wave heights ensuring that extreme events are prioritized in the parameter estimation procedure. Alternative choices that prioritize high wave heights could be, for example, linearly increasing weights, $w_i=x_i / \sum x_i$, or cubically increasing weights, $w_i = x_i^3 / \sum x_i^3$. We briefly tested these alternatives and, based on visual inspection of the estimated distributions, decided to weight errors quadratically. The outlined estimation method was implemented in Matlab. The code is open source (MIT license) and available at \url{https://github.com/ahaselsteiner/exponentiated-weibull}. In the appendix we describe the used mathematics and algorithms to solve Equation \ref{eq:argmin-exponentiated}.
\par 
To assess the uncertainty of the estimated parameters, we used bootstrapping with replacement (see, for example, \cite{Efron1993}). We estimated standard errors based on 100 bootstrap samples.
\subsection{Goodness of fit assessment}
We assessed each model's goodness of fit by computing the mean absolute error and by comparing each model's predicted 1-year return value with the empirical 1-year return value.  Both assessments are first performed with the original datasets ($A, B, C, ...$) and then with the retained datsets ($A_r, B_r, C_r, ...$).
\par 
Mean absolute error, $\bar{e}$, was first computed for the whole dataset:
\begin{equation}
    \bar{e} = \dfrac{\sum_{i=1}^n (x_i - \hat{x_i})}{n},
    \label{eq:overall-mae}
\end{equation}
Then, to assess the goodness of fit at high quantiles, we computed mean absolute error for quantiles with $p_i>0.99$ (`the tail') and for quantiles with $p_i>0.999$ (`the very tail'). The two errors, $\bar{e}_{0.99}$ and $\bar{e}_{0.999}$ read 
\begin{equation}
    \bar{e}_{[p_i]} = \dfrac{\sum_{i=j}^n( x_i - \hat{x_i})}{n-j+1}
    \label{eq:tail-mae}
\end{equation}
where $j$ is the index of the first empirical quantile whose $p_i$-value is above the threshold of 0.99 or 0.999.
\par 
For the assessment of the predicted 1-year return value, we computed the normalized return value, $H_{s1}^*$, by dividing the predicted return value, $\hat{H}_{s1}$, by the empirical return value, $H_{s1}$:
\begin{equation}
    H_{s1}^* = \dfrac{\hat{H}_{s1}}{H_{s1}}
\end{equation}
where $H_{s1}$ is the smallest empirical quantile whose probability, $p_i$, is greater than $(1 - p_e)$, with $p_e$ being the probability of exceedance, $p_e = 1 / (365.25 \times 24)$. For consistency, $\hat{H}_{s1}$ is computed using this empirical $p_i$-value too, instead of the exact value, which is $1-p_e$. Then $H_{s1}^*=1$ represents perfect agreement, $H_{s1}^*<1$ a too low prediction and $H_{s1}^*>1$ a too high prediction.
\section{Results}
\subsection{Estimated parameters and visual assessment}
\begin{figure*}
    \centering
    \includegraphics[width=\textwidth]{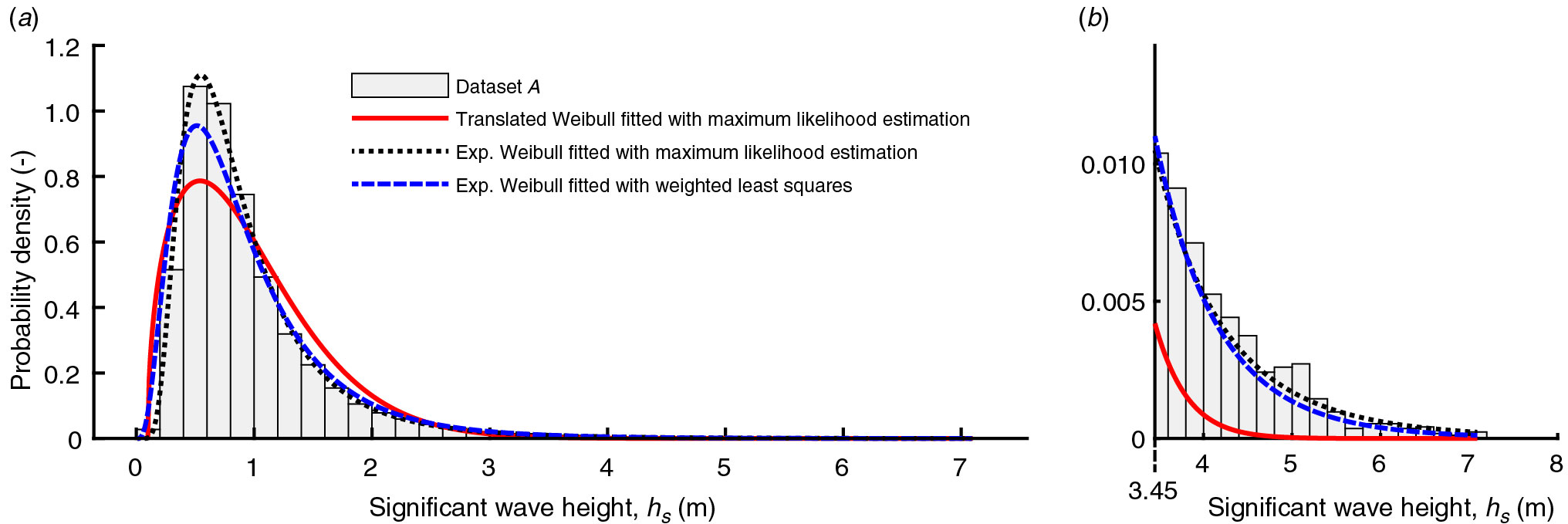}
    \caption{\textbf{Model fit between dataset A and the three considered models.} (\textit{a}) Complete range of the distribution. (\textit{b}) Tail of the distribution ($p_i > 0.99)$. The three models fit decently at the distribution's body, but at the tail the translated Weibull distribution underestimates the observed probability density. This behavior is present in all six datasets.}
    \label{fig:dataset-a}
\end{figure*}

\begin{figure*}
    \centering
    \includegraphics{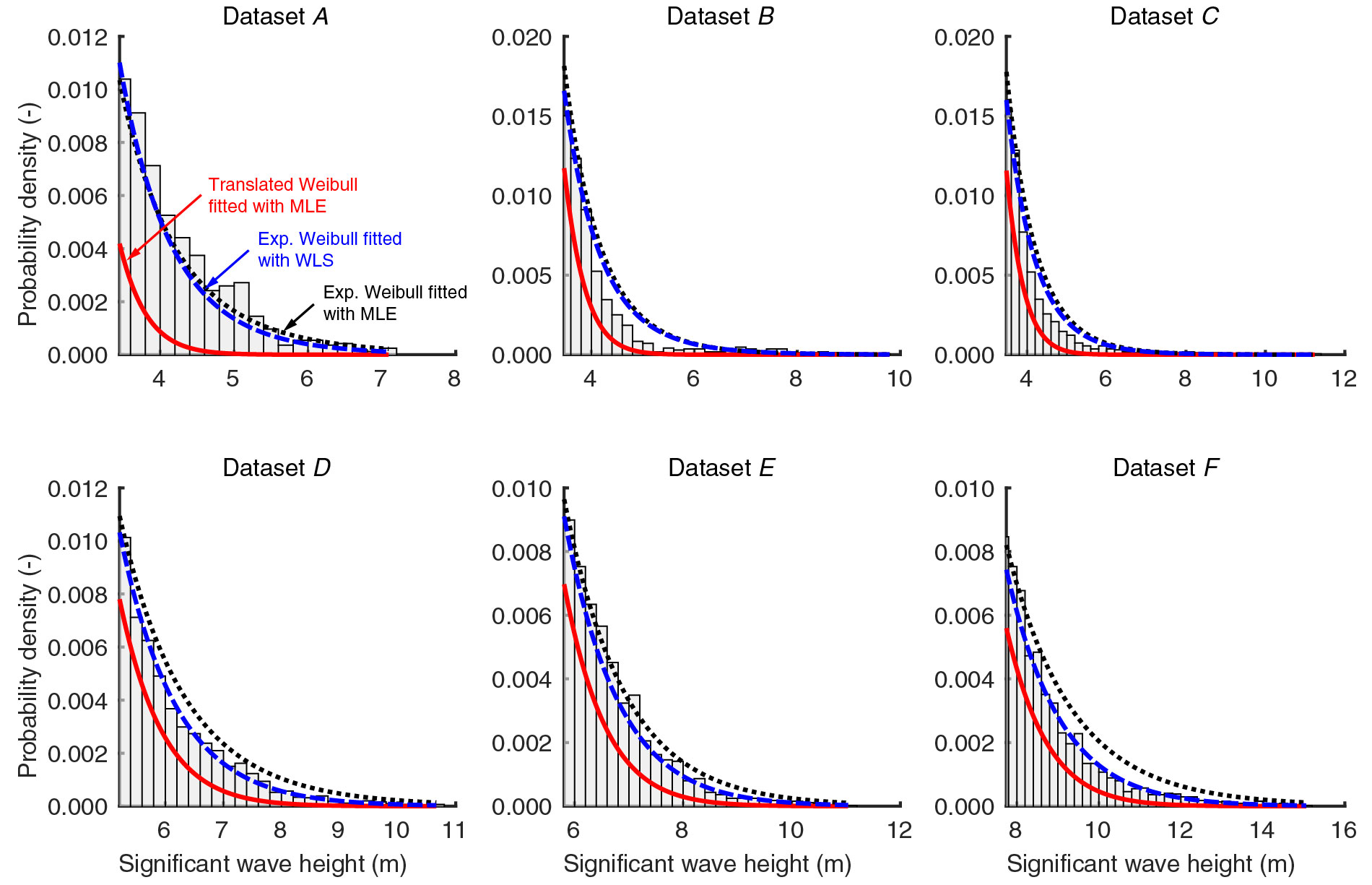}
    \caption{\textbf{Tail plots of all data sets ($p_i > 0.99$).} The translated Weibull distributions predict too low probability densities in all datasets. In dataset $D$ and $F$, the MLE-fitted exponentiated Weibull distributions predict too high probability densities.}
    \label{fig:tail-plots}
\end{figure*}

\begin{figure*}
    \centering
    \includegraphics[width=\textwidth]{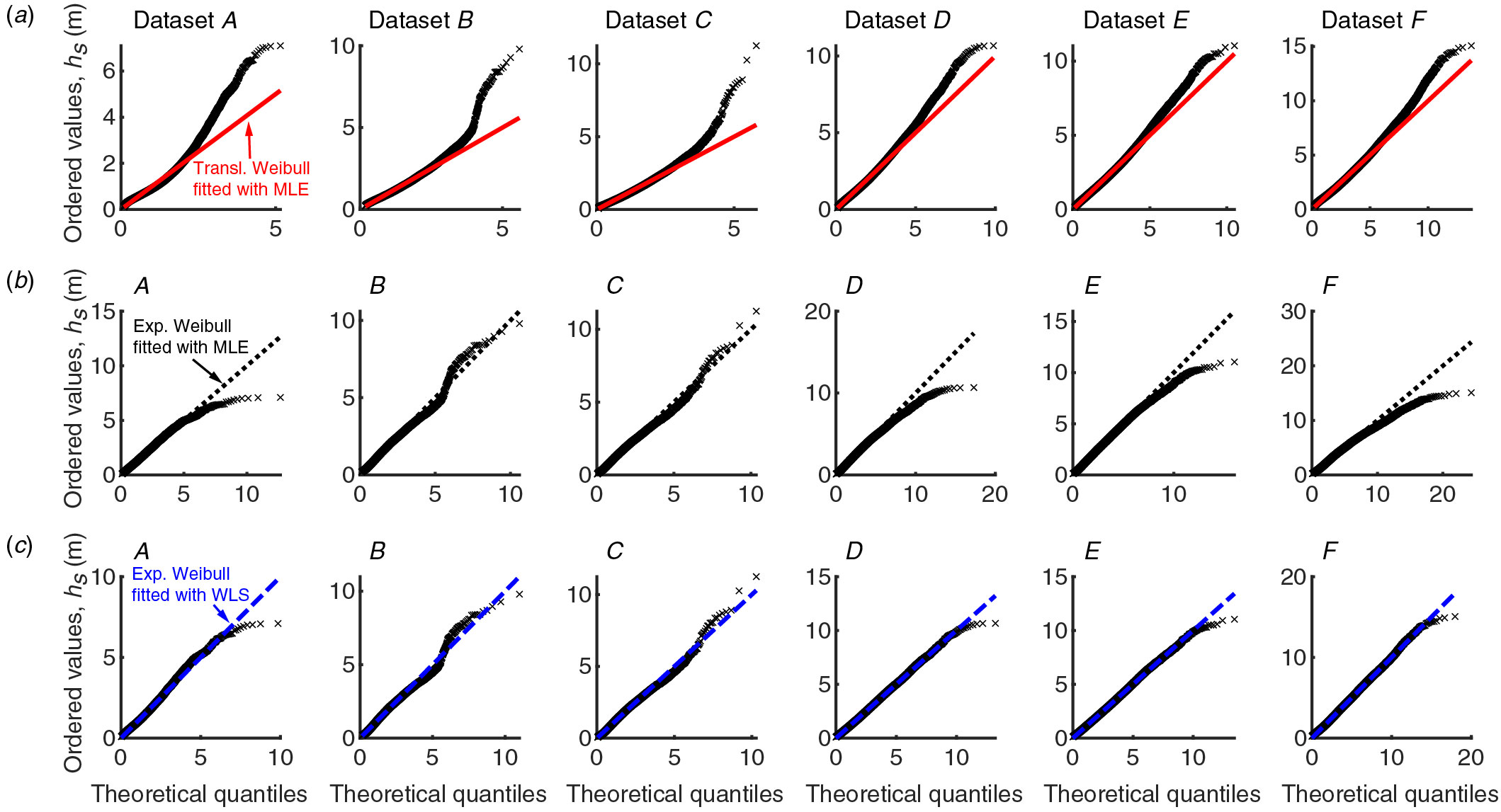}
    \caption{\textbf{QQ-plots of all data sets.} (\textit{a}) Translated Weibull distributions. (\textit{b}) Exponentiated Weibull distributions fitted with maximum likelihood estimation (MLE). (\textit{c}) Exponentiated Weibull distributions fitted with weighted least squares (WLS) estimation.}
    \label{fig:qq-plots-original-datasets}
\end{figure*}

The fitted translated Weibull distributions (Tab. \ref{tab:estimated-parameters-tranlated}) provide decent model fit at the bulk of the data, but fit poorly at the  tail. This is apparent both, in density plots (Fig. \ref{fig:dataset-a} \& \ref{fig:tail-plots}) and in QQ-plots (Fig. \ref{fig:qq-plots-original-datasets}). In all datasets, the translated Weibull distribution predicts too low probability densities in the tail ($p_i>0.99$; Fig. \ref{fig:tail-plots}) and consequently also too low quantiles in the tail (Fig. \ref{fig:qq-plots-original-datasets}\textit{a}). 
\par 
Density plots suggest that the fitted exponentiated Weibull distributions (Tab. \ref{tab:estimated-parameters-exponentiated}) provide good model fit at both, the body and the tail (Fig. \ref{fig:dataset-a} \& \ref{fig:tail-plots}). At the  tails, the densities of the MLE-fitted exponentiated Weibull distributions provide better model fit than the translated Weibull distributions. However, for dataset $D$ and $F$ the MLE-fitted exponentiated Weibull distribution predicts too high densities. The densities of the WLS-fitted distributions seem to better fit at these datasets. Overall, at the tail, the WLS-fitted exponentiated distributions match the empirical density values the closest.
\par 
The QQ-plots show similar results as the density plots: The MLE-fitted exponentiated Weibull distributions match the data better than the translated Weibull distributions at high quantiles  (Fig. \ref{fig:qq-plots-original-datasets}). However, at four datasets they predict too high values (datasets $A, D, E$ \& $F$). The WLS-fitted distributions provide good model fit over the complete range of the datasets. Only at the highest few observations deviations between the ordered values and the theoretical quantiles are apparent.
\par 
In summary, the density plots and the QQ-plots suggest that the exponentiated Weibull distribution is a better global model for significant wave height than the translated Weibull distribution. However, the QQ-plots show that in some datasets the MLE-fitted exponentiated Weibull distributions predict too high wave heights at high quantiles. There, the WLS-fitted distributions represent an improvement over the MLE-fitted distributions.
\par 

\begin{table*}[]
    \centering
    \begin{tabular}{l l l l}
         Dataset & $\alpha$ (scale) & $\beta$ (shape) & $\gamma$ (location) \\
         \hline
        $A$ & 0.9445$\pm$0.0055 & 1.4818$\pm$0.0097 & 0.0981$\pm$0.0039\\
        $B$ & 1.1413$\pm$0.0118 & 1.5990$\pm$0.0140 & 0.1878$\pm$0.0030\\
        $C$ & 1.1645$\pm$0.0124 & 1.5562$\pm$0.0166 & 0.0566$\pm$0.0097\\
        $D$ & 1.5797$\pm$0.0032 & 1.4067$\pm$0.0029 & 0.1024$\pm$0.0014\\
        $E$ & 1.8608$\pm$0.0027 & 1.4925$\pm$0.0028 & 0.1222$\pm$0.0007\\
        $F$ & 2.5693$\pm$0.0059 & 1.5466$\pm$0.0046 & 0.2248$\pm$0.0008\\
    \end{tabular}
    \caption{\textbf{Estimated parameters of the translated Weibull distributions.} Values after the $\pm$-sign represent the bootstrap estimate of the standard error.}
    \label{tab:estimated-parameters-tranlated}
\end{table*}

\begin{table*}[]
    \centering
    \begin{tabular}{l l l l l}
         Dataset & Method & $\alpha$ (scale) & $\beta$ (shape) & $\delta$ (shape) \\
         \hline
         \multirow{2}{*}{$A$}
         & MLE & 0.0373$\pm$0.0041 & 0.4743$\pm$0.0094 & \hfill 46.6078$\pm$3.8433\\
         & WLS & 0.2069$\pm$0.0149 & 0.6844$\pm$0.0142 & \hfill 7.7863$\pm$0.6239\\ 
         \hline
         \multirow{2}{*}{$B$}
         & MLE & 0.1731$\pm$0.0077 & 0.6563$\pm$0.0085 & \hfill 17.3927$\pm$0.7582\\
         & WLS & 0.0988$\pm$0.0259 & 0.5835$\pm$0.0316 & \hfill 36.5747$\pm$9.7319\\
         \hline
         \multirow{2}{*}{$C$}
         & MLE & 0.3026$\pm$0.0085 & 0.7445$\pm$0.0077 & \hfill 6.4434$\pm$0.1762\\
         & WLS & 0.2269$\pm$0.0735 & 0.6973$\pm$0.0636 & \hfill 9.8461$\pm$4.2791\\
         \hline
          \multirow{2}{*}{$D$}
         & MLE & 0.4728$\pm$0.0072 & 0.7452$\pm$0.0042 & \hfill 5.1186$\pm$0.0743\\
         & WLS & 0.9801$\pm$0.0278 & 1.0077$\pm$0.0147 & \hfill 2.1787$\pm$0.0805\\
         \hline
         \multirow{2}{*}{$E$}
         & MLE & 0.7889$\pm$0.0098 & 0.8842$\pm$0.0052 & \hfill 3.7615$\pm$0.0513\\
         & WLS & 1.2387$\pm$0.0249 & 1.0991$\pm$0.0120 & \hfill 2.0867$\pm$0.0601\\
         \hline
         \multirow{2}{*}{$F$}
         & MLE & 0.7180$\pm$0.0117 & 0.7663$\pm$0.0046 & \hfill 6.5994$\pm$0.1112\\
         & WLS & 1.6237$\pm$0.0387 & 1.0941$\pm$0.0141 & \hfill 2.4034$\pm$0.0824\\
    \end{tabular}
    \caption{\textbf{Estimated parameters of the exponentiated Weibull distributions.} Parameters are estimated either using maximum likelihood estimation (MLE) or weighted least squares (WLS).  Values after the $\pm$-sign represent the bootstrap estimate of the standard error.}
    \label{tab:estimated-parameters-exponentiated}
\end{table*}

\subsection{Quantitative assessment}
\begin{figure*}
    \centering
    \includegraphics[width=\textwidth]{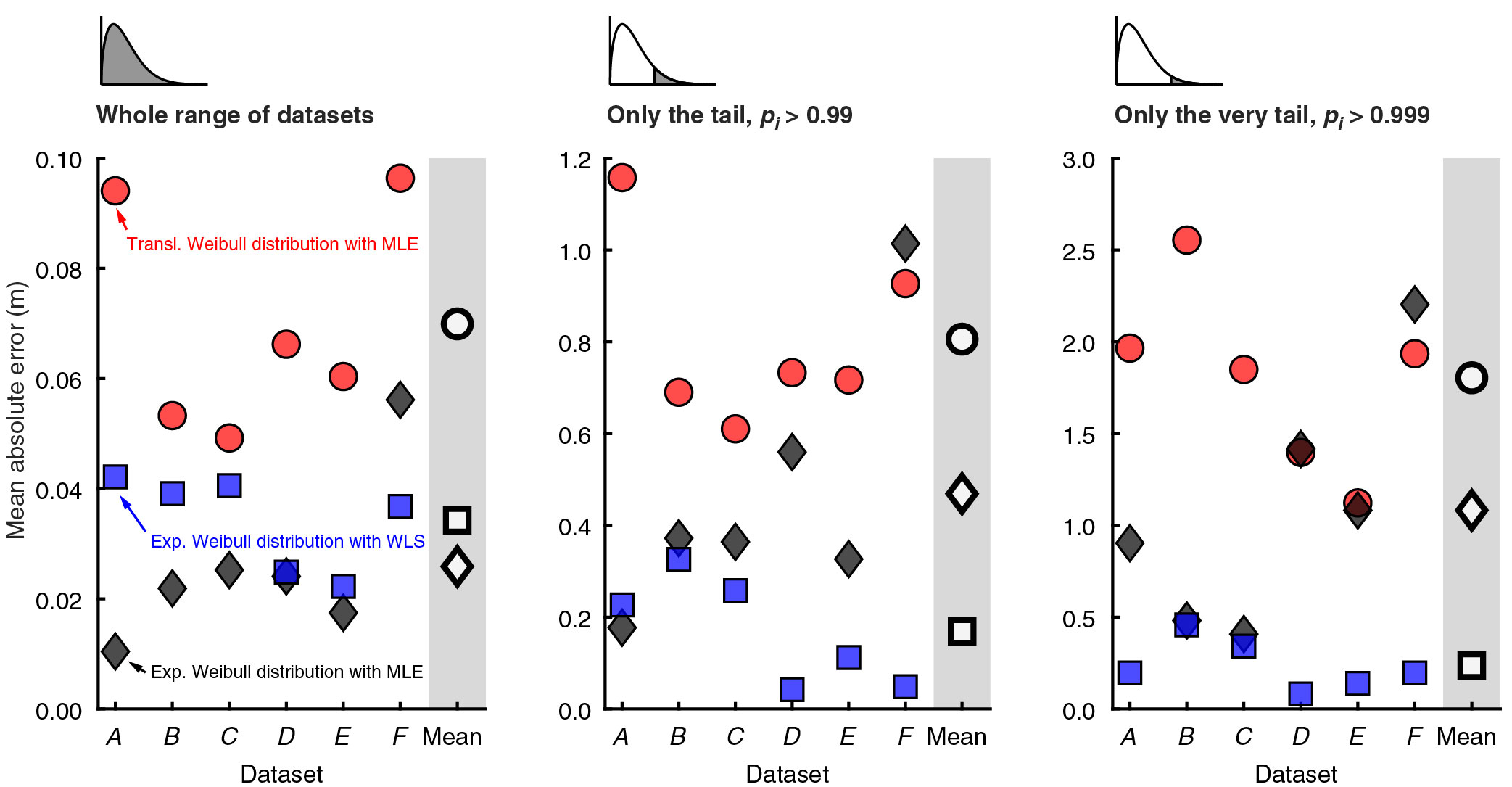}
    \caption{\textbf{Mean absolute error of the predicted significant wave height for different parts of the distribution.} Mean absolute error is a measure for the goodness of fit. It is calculated from the deviations between sample quantiles and predicted quantiles (see Equations \ref{eq:overall-mae} and \ref{eq:tail-mae}). Circles = translated Weibull distributions, diamonds = MLE-fitted exponentiated Weibull distributions, squares = WLS-fitted exponentiated Weibull distributions.}
    \label{fig:mae-original-datasets}
\end{figure*}

The two models of the exponentiated Weibull distribution provide the best fit in terms of mean absolute error (Fig. \ref{fig:mae-original-datasets}). When the whole range of the datasets is considered, the MLE-fitted distributions have the lowest mean absolute error in five of six datasets. In the tail ($p_i>0.99$) the WLS-fitted exponentiated Weibull distributions have the lowest mean absolute errors in five of six datsets. In the very tail ($p_i>0.999$) the WLS-fitted exponentiated Weibull distributions have the lowest errors in all datasets. There, the averaged mean absolute errors of the three models are $0.24\pm0.14$\,m, $1.08\pm0.67$\,m and $1.80\pm0.50$\,m (WLS-fitted exponentiated Weibull distribution, MLE-fitted exponentiated Weibull distribution and translated Weibull distribution, respectively;  $\bar{e}_{0.999}$-values are averaged over the six datasets, $N=6$; values after the $\pm$-sign represent standard deviations).
\par 
\begin{figure}
    \centering
    \includegraphics{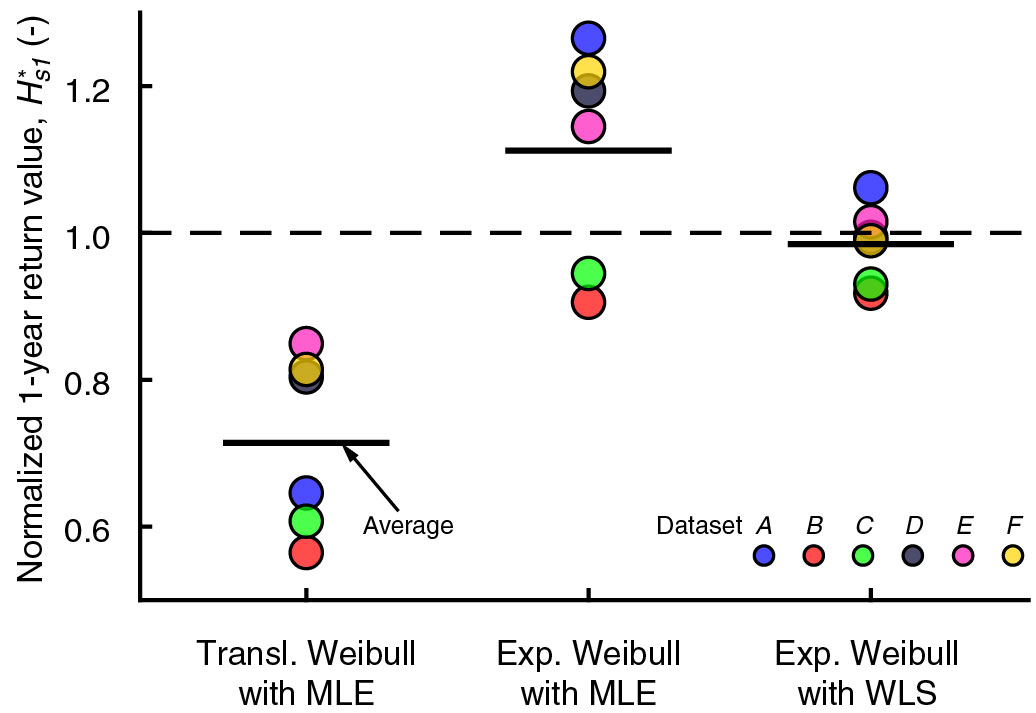}
    \caption{\textbf{Comparison of the 1-year return values predicted by the three considered models.} Normalized return values are calculated by dividing each model's return value by the empirical return value such that a value of 1 describes perfect agreement. The average of the normalized 1-year return values is too low for the translated Weibull distribution and too high for the MLE-fitted exponentiated Weibull distribution. The  1-year return values of the WLS-fitted exponentiated Weibull distribution agrees best with the empirical return values.}
    \label{fig:1-year-value-original}
\end{figure}

The empirical return values are best predicted by the WLS-fitted exponentiated Weibull distribution (Fig. \ref{fig:1-year-value-original}). Its averaged normalized 1-year return value is 0.985, its standard deviation 0.054 ($N=6$). The translated Weibull distribution predicts too low 1-year return values in all datasets ($H_{s1}^*=0.714\pm0.122$, $N=6$) and the MLE-fitted exponentiated Weibull distribution predicts too high return values in four datasets ($H_{s1}^*=1.112\pm0.151$, $N=6$).
\par
\begin{figure}
    \centering
    \includegraphics{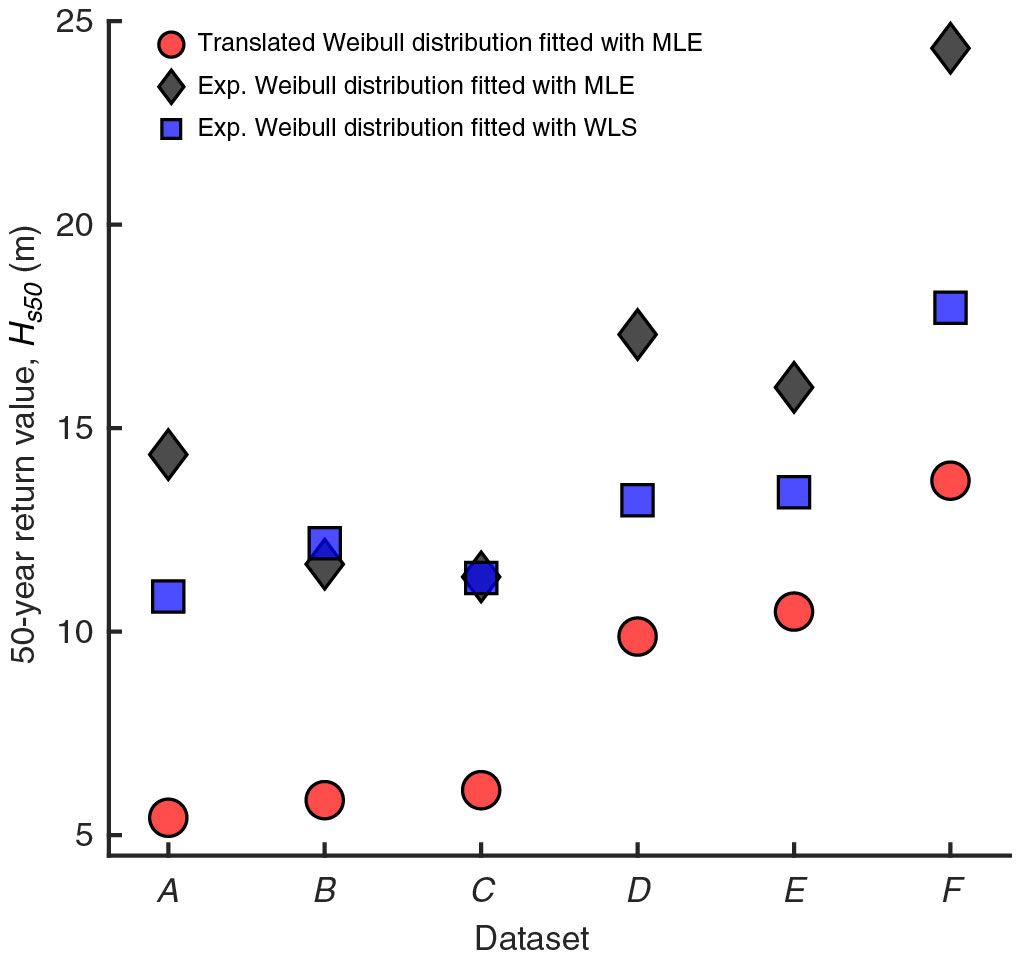}
    \caption{\textbf{Predictions of the 50-year return values.} The translated Weibull distribution predicts the lowest return value in all considered datasets.}
    \label{fig:50-year-return-values}
\end{figure}

 
The three type of models lead to big differences when 50-year return values are predicted (Fig. \ref{fig:50-year-return-values}). For example, for dataset $A$ the translated Weibull distribution predicts $H_{s50}=5.43$\,m, the MLE-fitted exponentiated Weibull distribution predicts $H_{s50}=14.35$\,m and the WLS-fitted distribution predicts $H_{s50}=10.86$\,m. For comparison, in dataset $A$, which covers only a duration of 10 years, the highest measured $H_s$-value is 7.10\,m.

\begin{figure*}
    \centering
    \includegraphics[width=\textwidth]{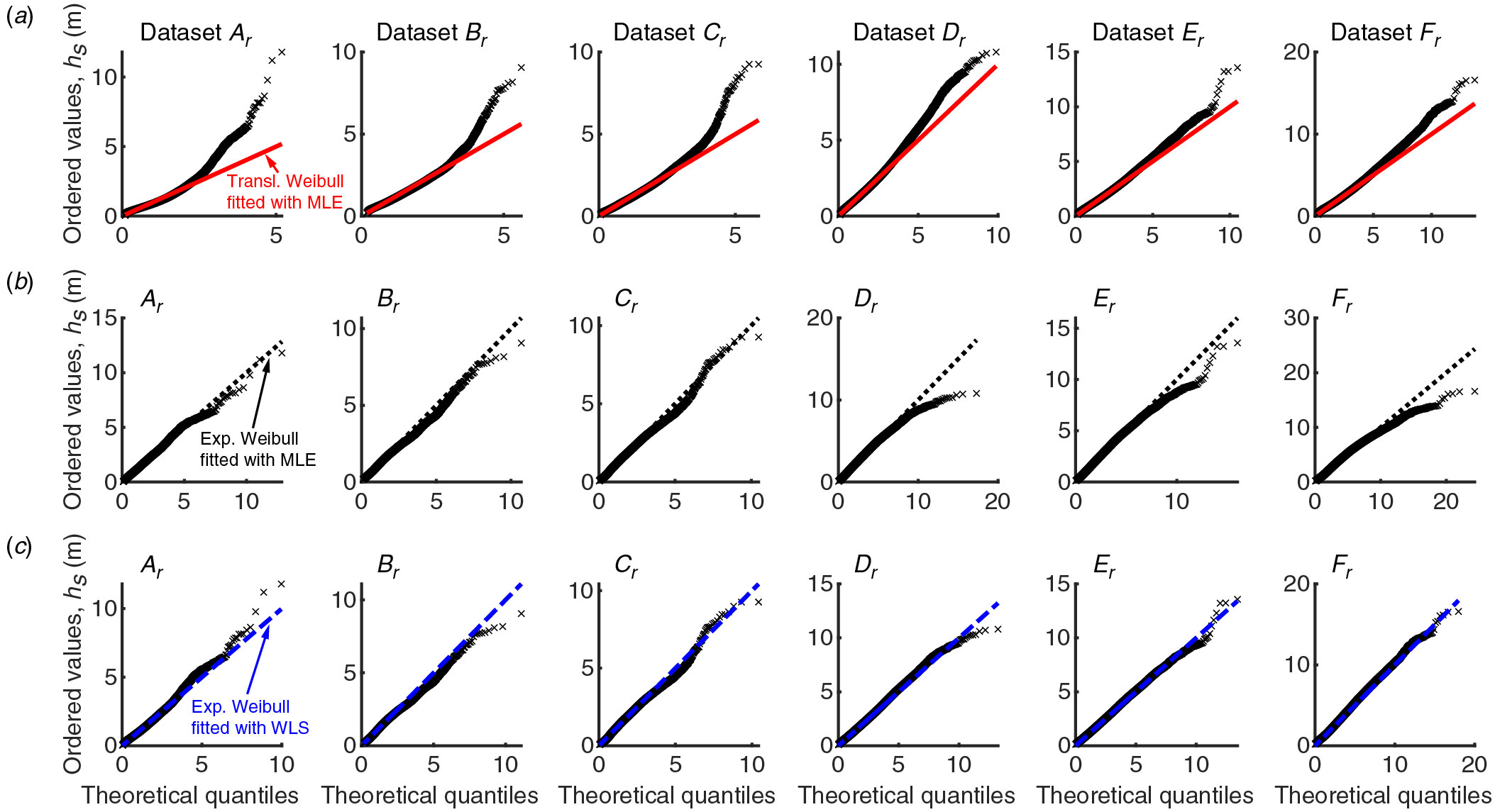}
    \caption{\textbf{QQ-plots of the fitted distributions and the retained datasets.} (\textit{a}) Translated Weibull distributions. (\textit{b}) Exponentiated Weibull distributions fitted with maximum likelihood estimation. (\textit{c}) Exponentiated Weibull distributions fitted with weighted least squares estimation.}
    \label{fig:qq-plots-retained-data}
\end{figure*}

As a possibly more direct assessment of how well the fitted distributions predict future wave heights, we used some retained parts of the used data sources. The results obtained with these retained datasets are similar to the results with the original datasets: QQ-plots show that the WLS-fitted exponentiated Weibull distributions provide good model fit at low, medium and high quantiles (Fig. \ref{fig:qq-plots-retained-data}). Further, the translated Weibull distributions predict too low wave heights at high quantiles and the MLE-fitted exponentiated Weibull distributions sometimes predict too high wave heights at very high quantiles.
\par
\begin{figure*}
    \centering
    \includegraphics[width=\textwidth]{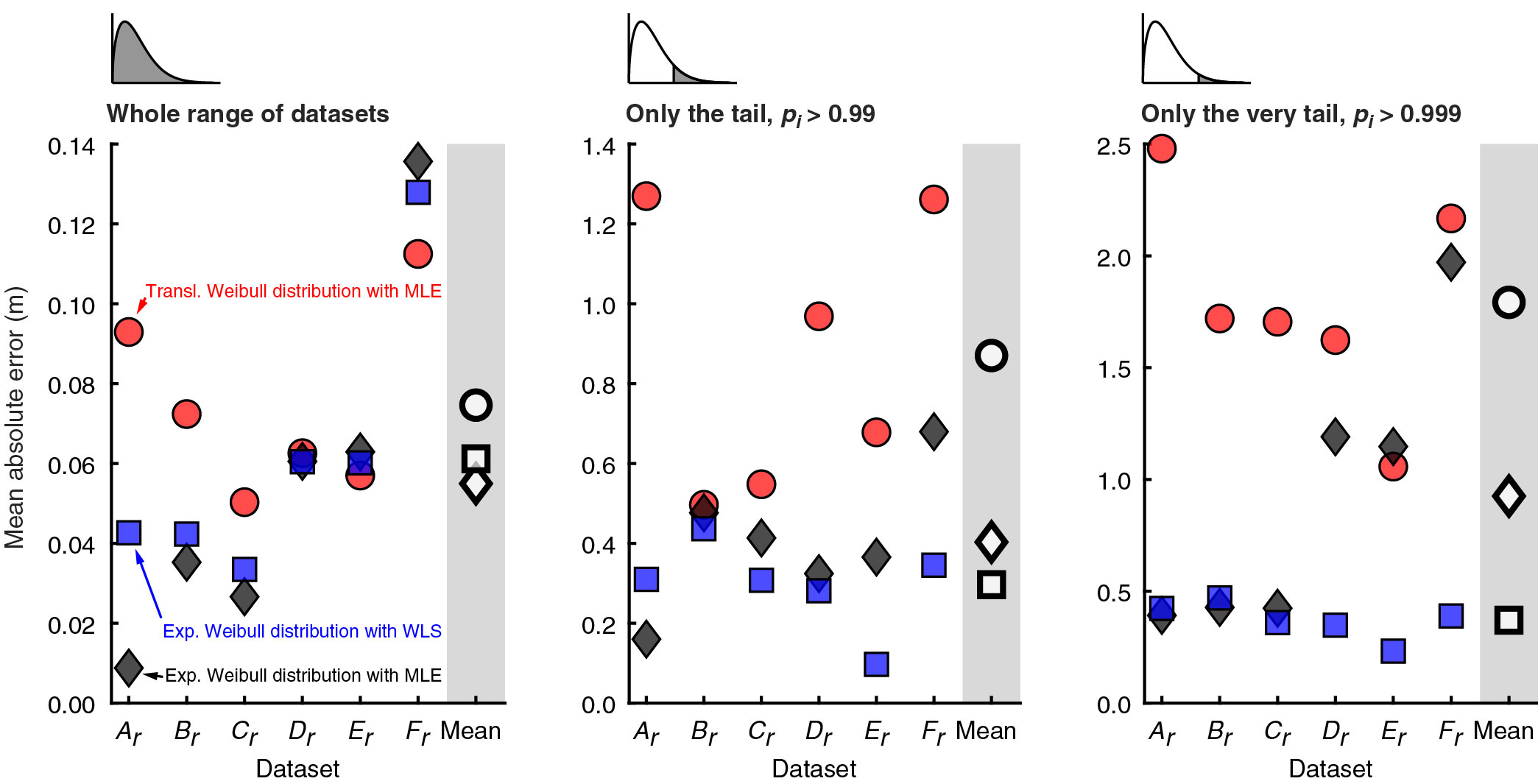}
    \caption{\textbf{Mean absolute error between the fitted distributions and the retained datasets.} In the very tail ($p_i>0.999$) the WLS-fitted exponentiated Weibull distribution has the lowest error. Circles = translated Weibull distributions, diamonds = MLE-fitted exponentiated Weibull distributions, squares = WLS-fitted exponentiated Weibull distributions.}
    \label{fig:mae-retained-data}
\end{figure*}

Among all models and datasets, the overall mean absolute error is between 0.01 and 0.14\,m and no model is best or worst among all datasets (Fig. \ref{fig:mae-retained-data}). In the very tails ($p_i>0.999$), the WLS-fitted exponentiated Weibull distributions have the lowest averaged mean absolute error, $0.37\pm0.08$\,m ($N=6$). The averaged mean absolute error of the MLE-fitted exponentiated Weibull distributions is $0.93\pm0.63$\,m ($N=6$) and the averaged mean absolute error of the translated Weibull distributions is $1.79\pm0.49$\,m ($N=6$; both also for $p_i>0.999$).
\par 

\begin{figure}
    \centering
    \includegraphics{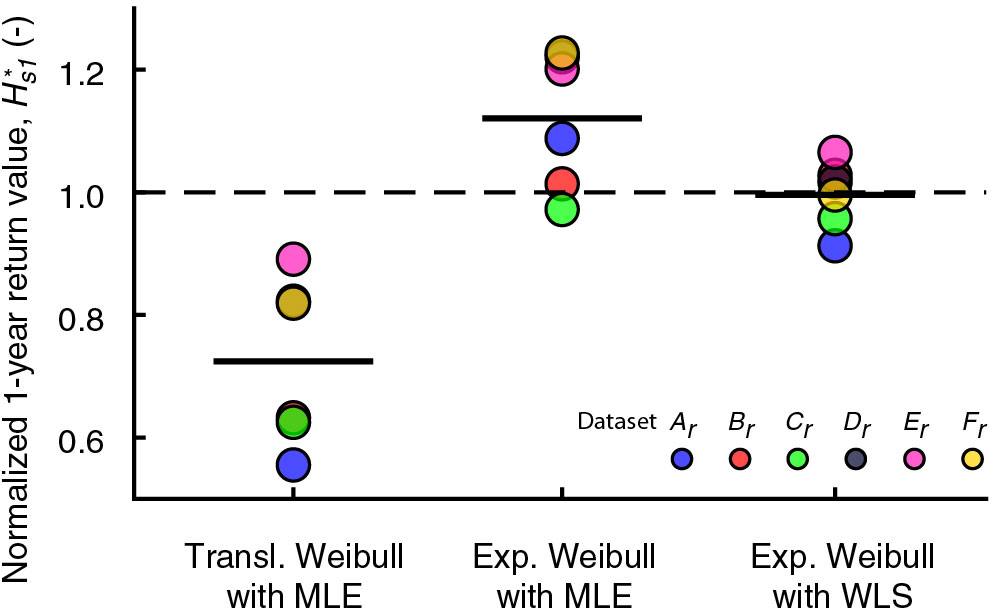}
    \caption{\textbf{Comparison of the predicted 1-year return values and the return values of the retained datasets.} Normalized return values are calculated by dividing each model's return value by the empirical return value such that a value of 1 describes perfect agreement.}
    \label{fig:1-year-return-value-retained}
\end{figure}

When the predicted 1-year return values are compared with the empirical 1-year return values of the retained datasets, the results are similar as in the comparison with the values of the original datasets (Fig. \ref{fig:1-year-return-value-retained}): The translated Weibull distributions predict too low wave heights ($H_{s1}^*=0.724\pm0.137$, $N=6$), the MLE-fitted exponentiated Weibull distributions predict mostly too high wave heights ($H_{s1}^*=1.120\pm0.112$, $N=6$) and the WLS-fitted distributions match the empirical return values best ($H_{s1}^*=0.996\pm0.054$, $N=6$).
\par 
In summary, the quantitative assessment showed that the three models have overall mean errors in the same order of magnitude. Among all datasets, all models and both in-sample prediction (datasets $A, B, ...$) and out-of-sample prediction (datasets $A_r, B_r, ...$), the overall mean error was between 0.01 and 0.14\,m. Overall mean error and QQ-plots suggest that the three models perform relatively similar for typical $H_s$-values. In the very tail ($p_i > 0.999$), however, a clear ranking of model performance is appararent: The WLS-fitted exponentiated Weibull distribution has the lowest averaged mean absolute error (0.24\,m in-sample prediction, 0.37\,m out-of-sample prediction) and the translated Weibull distribution has the highest error (1.80\,m in-sample, 1.79\,m out-of-sample). 
\section{Discussion}
\subsection{Comparison between the tested and other models}
Our analysis suggests that the exponentiated Weibull distribution is a better model for significant wave height than the translated Weibull distribution. Especially at the tail, the exponentiated Weibull distribution matches the empirical data better. However, when the distribution is fitted using maximum likelihood estimation, the estimated parameters are mainly driven by the bulk of the data and not by the data in the very tail. Consequently, in our analysis, considerable errors remained for high quantiles such as the 1-year return value. In marine structural design, high quantiles of significant wave height are especially important. Thus, to improve model fit at the tail, we estimated the distribution's parameters by minimizing the sum of the weighted squared errors between data and model. To prioritize high values, we weighted the errors based on the squared wave height value.
\par 
The strongest possible prioritization of observations of high wave heights would be to ignore observations up to a particular threshold. In this case one would fit a distribution solely to the tail. To create a global model, a second distribution could be fitted to the bulk of the data. The combination of these two distributions creates a `two-part model', which would serve as a global model. In such a model, the tail could be modelled with a generalized Pareto distribution and the body could be modelled, for example, with a Weibull distribution. The generalized Pareto distribution is often used to model the tails of other distributions and has been considered for wave heights (see, for example, \cite{Thompson2009, Teixeira2018}) 
We expect that a two-part model could estimate the tail even better than the model that we proposed, however, that is expected for a model that has more parameters (2-3 parameter for the tail and 2-3 parameter for the body). Further, two-part models either have a discontinuity in the PDF at the transition between the two distributions or they enforce continuity as another boundary condition, which might weaken the goodness of fit to the data (for a review on extreme value threshold estimation, see Scarrott and MacDonald \cite{scarrott:2012}).
\par 
Other models that have been proposed as global models for the significant wave height are the log-normal distribution \cite{Jasper1956}, the 2-parameter Weibull distribution \cite{Battjes1972}, the generalized gamma distribution \cite{Ochi1992}, the 3-parameter beta distribution of the second kind \cite{Ferreira1999}, the `Ochi distribution' \cite{Ochi1980} and the `Lonowe distribution' \cite{Haver1985}. The 2-parameter Weibull distribution and the log-normal distribution have only two parameters and -- to justify that we propose to use a distribution with three parameters -- should fit clearly worse to the data. These 2-parameter distributions have been considered to be insufficient by other authors (see, for example, \cite{Ochi1980, li:2015}) and a brief inspection we performed suggested the same thing for the datasets of this study. 
\par
\begin{figure*}
    \centering
    \includegraphics{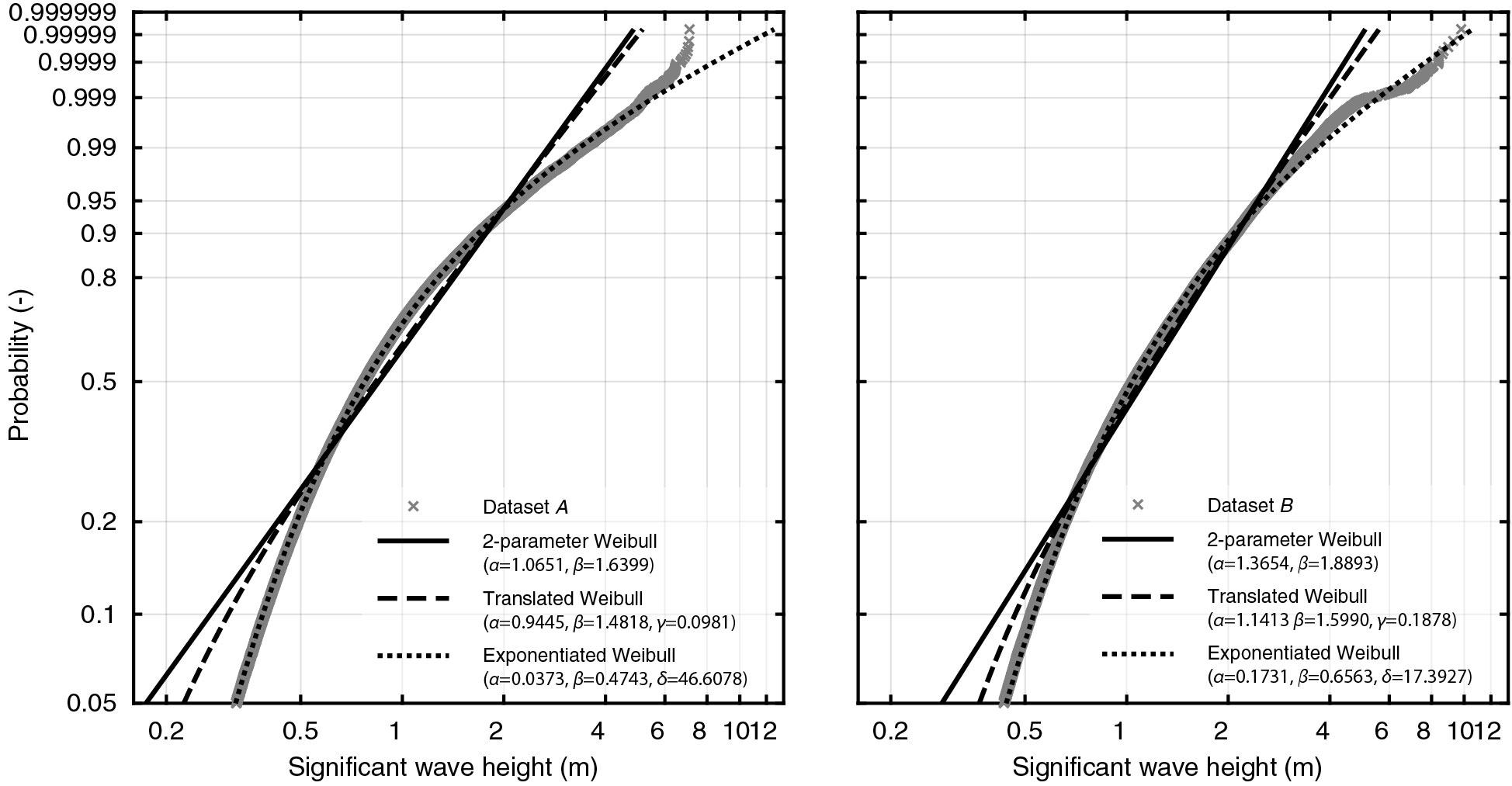}
    \caption{\textbf{Weibull probability plot of datasets \textit{A} and \textit{B}.} The shape parameter $\delta$ enables the exponentiated Weibull distribution to prescribe a curved line that is similar to the empirical data. The parameters of the shown distributions were fitted using maximum likelihood estimation.}
    \label{fig:weibull-paper}
\end{figure*}

 
To understand why the exponentiated Weibull distribution provides much better model fit than the common 2-parameter Weibull distribution, plotting the data on Weibull paper is illuminating (Fig. \ref{fig:weibull-paper}). On Weibull paper, the wave data does not follow a straight line, but a continuously bending curve. The exponentiated Weibull distribution's second shape parameter, $\delta$, enables the distribution to follow this bend: $\delta>1$ will lead to a curve that bends to the right and  $\delta<1$ will lead to a curve that bends to the left. The translated Weibull distribution's location parameter also leads to a slight bend when plotted on Weibull probability paper. However, its location parameter does not control the shape directly and consequently the translated Weibull distribution cannot match the empirical wave data to a similar degree as the exponentiated Weibull distribution.

\par 
The Ochi distribution and the Lonowe distribution have more parameters than the exponentiated Weibull distribution: The Ochi distribution has four parameters and the Lonowe distribution has four or five (depending on whether one counts the threshold between the Weibull-part of the model and the lognormal-part of the model as a parameter or not). Since we consider the performance of the exponentiated Weibull distribution as sufficiently good, we did not consider these more complex distributions in this study.
\par 
The generalized gamma distribution \cite{Ochi1992}, and the 3-parameter beta distribution of the second kind \cite{Ferreira1999}, however, have similar model complexity in terms of number of parameters. Thus, we tested these distributions by fitting them to the six datasets using maximum likelihood estimation and by computing the overall mean absolute error (details are provided in the appendix). The generalized gamma distributions and the 3-parameter beta distributions had errors of 0.0317$\pm$0.0203\,m and 0.0294$\pm$0.0177\,m, respectively ($N=6$). These averaged errors are higher than the averaged error of the MLE-fitted exponentiated Weibull distributions (0.0259$\pm$0.0158\,m, $N=6$), however, in the same order of magnitude. In two datasets, the exponentiated Weibull distribution had the lowest error and in four datasets, the generalized gamma distribution had the lowest error. Thus these three models seem to perform roughly equally well. Future research based on more datasets could help to find out more detailed differences between these three distributions.
\par
In summary, among the variety of possible global models for significant wave height, the exponentiated Weibull distribution represents a good compromise between model complexity and model accuracy: It performs better than the currently most used model -- the translated Weibull distribution -- without increasing model complexity. The good performance of the exponentiated Weibull distribution can be explained by its second shape parameter, $\delta$, which allows the distribution to represent a bending curve when plotted on Weibull paper.

\subsection{Implications on \replaced{design loads}{design load cases}}
The current international standards that regulate the design of fixed and floating offshore wind turbines \cite{IEC61400-3-1:2019-04, IECTS61400-3-2:2019-04} require designers to estimate the 1-year and 50-year return values of the significant wave height. \deleted{The 1-year return value, $H_{s1}$, is required for the design load cases (DLCs) 6.3, 7.1, 8.2, while the 50-year return value $H_{s50}$ is required for DLC 6.2 \cite{IEC61400-3-1:2019-04}.} \added{These return values are used in so-called design load cases (DLCs). A design load case describes an operating condition of a wind turbine, together with the environmental conditions during the particular operating condition. They are used to check whether a wind turbine design preserves structural integrity under all future environmental and operating conditions that can reasonable be expected. DLCs are developed and maintained by standardization organizations. The International Electrotechnical Commission's standard IEC 61400-3 \cite{IEC61400-3-1:2019-04} is widely used by turbine manufacturers, certifying organizations and by academics who study wind turbine design (see, for example, \cite{Robertson2011,Morato2017}). The standard abbreviates design load cases with numbers. Estimating the 1-year wave height return value, $H_{s1}$, is required for the design load cases 6.3, 7.1 and 8.2, while the 50-year return value, $H_{s50}$, is required for DLC 6.2 \cite{IEC61400-3-1:2019-04}. In each of these design load cases, the estimated wave height return value determines a design load that is used to evaluate structural integrity. Consequently, which model is used to describe the distribution of significant wave height influences wind turbine design via these design load cases.}
\par 
Our results show that the current common technique of fitting a translated Weibull distribution to significant wave height data using maximum likelihood estimation underestimates the 1-year return value strongly. At the six tested sites, the 1-year return value is underestimated on average by about 30\%. In some cases the underestimation is even more severe. For example, in dataset $A$, the empirical 1-year return value is about 6.7\,m, but the fitted translated Weibull distribution predicts a wave height of only 4.3\,m.
\par 
In structural design, uncertainties are partly taken care of with safety factors, which are multiplied to design loads (for details, see, for example, \cite[][pp. 66-68]{IEC61400-3-1:2019-04}). The normal safety factor for offshore wind turbines is 1.35 \cite{IEC61400-3-1:2019-04}, which is of similar magnitude as the typical error when $H_{s1}$ is estimated based on a fitted translated Weibull distribution. This suggests that the found errors \replaced{can be critical for the safety of a wind turbine design, especially if a turbine is particularly wave-sensitive (for a discussion on wave-sensitive turbine design, see, for example, \cite{Velarde2019})}{are critical for the safety of wind turbine designs}. The differences in the estimated 50-year return values are potentially even bigger: For example, for dataset $B$, the translated Weibull distribution predicts a 50-year return value of about 6\,m, while the WLS-fitted exponentiated Weibull distribution predicts a return value of about 12\,m.
\par
\begin{figure*}
    \centering
    \includegraphics[width=\textwidth]{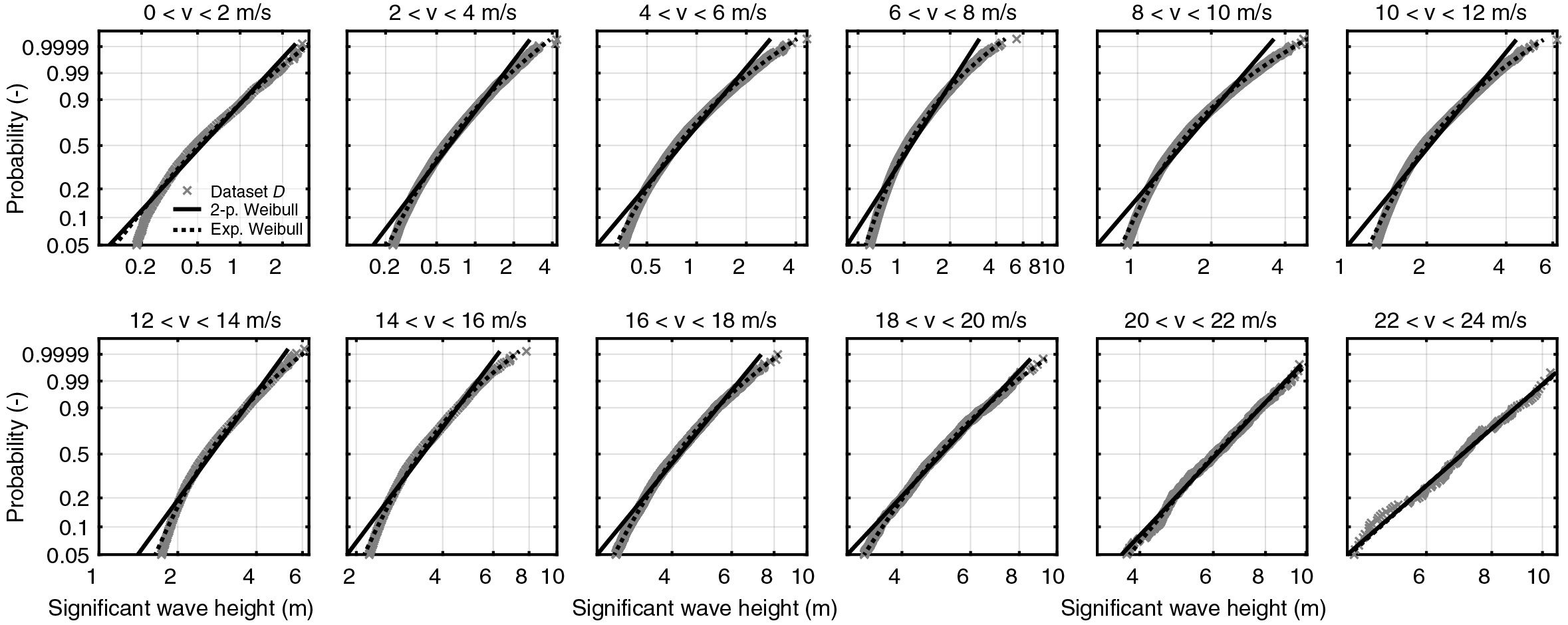}
    \caption{\textbf{Weibull probability plots of conditional wave height distributions.} Dataset $D$ was sorted into wind speed intervals ($v$ = wind speed). Due to the shape parameter, $\delta$, the exponentiated Weibull distributions (dotted lines) can follow the curved lines of the empirical data. The shown exponentiated Weibull distributions were fitted using weighted least squares estimation.}
    \label{fig:conditonal-hs}
\end{figure*}

 
Besides the unconditonal distribution of $H_s$, the offshore wind standard IEC 61400-3-1 \cite{IEC61400-3-1:2019-04} requires designers to estimate joint 50-year extremes of wind speed and wave height. In the standard's DLC 1.6 designers need to estimate the conditional wave height distribution for a given wind speed, that is $F(h_s|v)$. The standard does not prescribe which distribution should be assumed for $F(h_s|v)$, however, researchers usually assume that the conditional wave height follows a 2-parameter Weibull distribution (see, for example, \cite{li:2015,Liu2019}). We tested how the 2-parameter Weibull distribution and the exponentiated Weibull distribution fit to conditional wave height data, using dataset $D$ (the hindcast coastDat-2 also contains hourly wind data). Visual inspecton of Weibull probability paper plots suggest that the exponentiated Weibull distribution fits better (Fig. \ref{fig:conditonal-hs}). Future research could investigate how much better the exponentiated Weibull distribution performs across multiple datasets and which expressions could be used to model the dependence functions of the parameters $\alpha$, $\beta$ and $\gamma$.
\section{Conclusions}
In this paper, we showed that the exponentiated Weibull distribution matches the empirical distribution of significant wave height data better than the commonly used translated Weibull distribution. Since the exponentiated Weibull distribution does not add complexity when compared with the translated Weibull distribution, we argue that it represents a better global model for significant wave height. In the six analyzed datasets, the translated Weibull distribution always predicted too low 1-year return values. When the exponentiated Weibull distribution was fitted using maximum likelihood estimation, it predicted too high 1-year return values in four of six cases. To improve its fit at the tail, we estimated its parameters by minimizing the weighted squared error between the model and the observations. The weights were chosen to quadratically increase with wave height. These WLS-fitted distributions showed good fit over the complete range of the datasets. Overall mean absolute error was in the order of 0.1\,m and at the very tails ($p_i>0.999$) mean absolute error was in the order of 0.5\,m. Based on our results, we argue that if data do not indicate otherwise, the exponentiated Weibull distribution should be fitted to wave data instead of the translated Weibull distribution.

\section*{Data availability and open source Matlab implementation}
\label{sec:data-and-matlab}
The complete analysis performed in this study and the creation of the presented figures can be reproduced by running the file \texttt{CreateAllFigures.m} that is available in the repository \url{https://github.com/ahaselsteiner/2019-paper-predicting-wave-heights}. This repository contains also all datasets – preprocessed and structured as they were used in this study. Alternatively, the raw data of this study can be downloaded from the NDBC website, \url{www.ndbc.noaa.gov}, and from the coastDat-2 repository,  doi: \href{https://doi.org/10.1594/WDCC/coastDat-2_WAM-North_Sea}{10.1594/WDCC/coastDat-2\_WAM–North\_Sea}.
\par
The considered distributions were implemented in custom Matlab code. We created a Matlab class for each distribution. These classes provide functions to assess the PDF, the CDF, the ICDF, to fit the distribution's parameters and to draw samples from the distribution. They are available at
\par 
\begin{footnotesize}
\begin{itemize}
    \item \url{https://github.com/ahaselsteiner/exponentiated-weibull}, 
    \item  \url{https://github.com/ahaselsteiner/translated-weibull},
    \item \url{https://github.com/ahaselsteiner/generalized-gamma} and
    \item \url{https://github.com/ahaselsteiner/beta-3p-second-kind}.
\end{itemize}
\end{footnotesize}
\section*{Acknowledgements}
We thank V. Vutov for many useful discussions and for constructive criticism of the manuscript. Thanks to M. Brink, M. Haselsteiner and A. Sander for constructive criticism of the manuscript.

\appendix

\section{WLS-based estimators}
\label{sec:wls-estimators}
Our estimation method of the three parameters of the exponentiated Weibull distribution, $\alpha$, $\beta$ and $\delta$ is based on the `Weibull paper linearization' that is commonly used for the 2-parameter Weibull distribution (see, for example, Scholz \cite{Scholz2008}). In the following, we will describe our method in detail.
\par 
The ICDF of the exponentiated Weibull distribution reads
\begin{equation}
    x = \alpha [- \log_e (1 - p^{1/\delta})]^{1/\beta}.
\end{equation}
Taking the logarithm with base 10 we get
\begin{equation}
    \log_{10}(x) = \log_{10}(\alpha) + \dfrac{1}{\beta} \log_{10}[-\log_e(1 - p ^{1/\delta})],
\end{equation}
which shows a linear relationship between $\log_{10}(x)$ and $\log_{10}[-\log_e(1 - p ^{1/\delta})]$.
\par 
Thus, if we write $\log_{10}(x)=x^*$, $\log_{10}(\alpha)=a$, $\dfrac{1}{\beta}=b$ and $\log_{10}[-\log_e(1 - p ^{1/\delta})]=p^*$ we get the simple expression
\begin{equation}
    x^* = a + b p^*.
\end{equation}
This linear relationship allows us to use standard linear regression techniques to estimate the parameters $a$ and $b$ and, with these parameters, the distribution's parameters $\alpha$ and $\beta$.
\par 
Here, we have chosen to minimize the weighted squared deviations between the observed and the predicted values. Let the function $Q$ express the sum of the weighted squared errors: 
\begin{equation}
    Q(a,b; \delta) = \sum_{i=1}^n w_i (x_i^* - \hat{x}^*_i)^2 = \sum_{i=1}^n w_i [x_i^* - (a + b p_i^*)]^2,
\end{equation}
where $p^*_i$ is the normalized $p_i$-value, 
\begin{equation}
    p_i^* = \log_{10}[-\log_e(1 - p_i ^{1/\delta})].
\end{equation}
We can find the WLS-estimators $\hat{a}$ and $\hat{b}$ by differentiating $Q(a,b)$ and finding its root: 
\begin{equation}
    \dfrac{\partial{Q}(a,b)}{\partial a} = -2\sum_{i=1}^n w_i [x_i^* - (a + b p_i^*)] = 0,
    \label{eq:dQ-da}
\end{equation}

\begin{equation}
    \dfrac{\partial{Q}(a,b)}{\partial b} = -2\sum_{i=1}^n w_i p_i^* [x_i^* - (a + b p_i)] = 0 .
    \label{eq:dQ-db}
\end{equation}
Solving for $a$ in Equation \ref{eq:dQ-da} leads to
\begin{equation}
    \hat{a} = \bar{x}^* - \hat{b}\bar{p}^*,
    \label{eq:a-hat}
\end{equation}
where $\bar{x}^* = \sum_{i=1}^n w_i x_i^*$ and $\bar{p}^* = \sum_{i=1}^n w_i p_i^*$.
Similarly, by solving for $b$ in Equation \ref{eq:dQ-db} and by using Equation \ref{eq:a-hat}, we can derive an expression for $\hat{b}$: 
\begin{equation}
    \hat{b} = \dfrac{\sum_{i=1}^n (w_i p_i^* x_i^*) - \bar{x}^*\bar{p}^*}{\sum_{i=1}^n(w_i p_i^{*2}) - \bar{p}^{*2}}
\end{equation}
With $\hat{a}$ and $\hat{b}$ we can calcuate $\hat{\alpha}$ and $\hat{\beta}$: 
\begin{equation}
    \hat{\alpha} = 10^{\hat{a}},
\end{equation}
\begin{equation}
    \hat{\beta} = 1 / \hat{b}.
\end{equation}
Thus, for any given $\delta$-value we can explicitly compute the WLS-estimators $\hat{\alpha}$ and $\hat{\beta}$.
\par 
We are, however, still missing an expression for the estimator $\hat{\delta}$. To derive this expression, let us define a function that returns the weighted squared error of an exponentiated Weibull distribution with a given parameter $\delta$ as \begin{equation}
    Q_\delta(\delta) = Q(\hat{a}, \hat{b}; \delta).
\end{equation}
The WLS-estimator $\hat{\delta}$ is the $\delta$-value that minimizes this function: 
\begin{equation}
    \hat{\delta} = \argminB_\delta[Q_\delta(\delta)].
    \label{eq:delta-hat}
\end{equation}
We did not try to find an analytical solution to Equation \ref{eq:delta-hat}. Instead, we used Matlab's function \texttt{fminsearch.m} to compute the minimum.
\par
\begin{figure}
    \centering
    \includegraphics{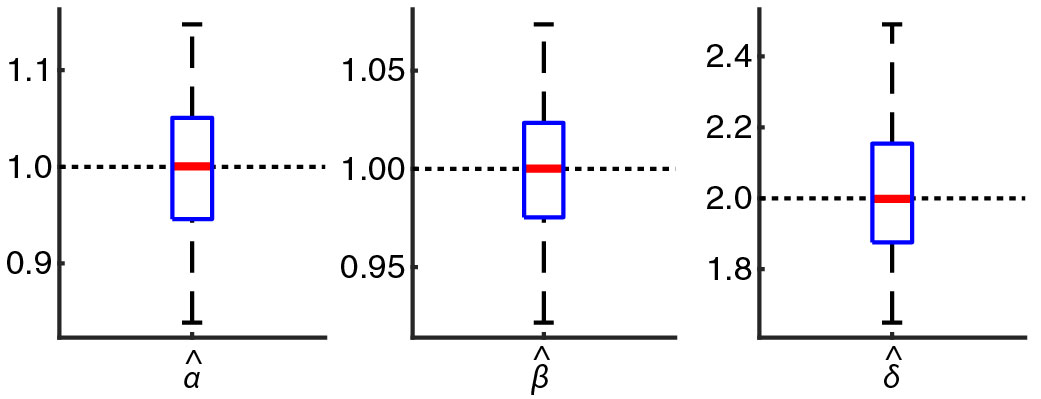}
    \caption{\textbf{Box plots of the estimated parameters of an exponentiated Weibull distribution.} The true distribution has the parameters $\alpha = 1$, $\beta = 1$ and $\delta = 2$. The thick line represents the median and the box the 25th and 75th percentile. 100 samples, each with 100,000 datapoints were used for the estimation.}
    \label{fig:evaluation-of-implementation}
\end{figure}

 
To evaluate whether the implemented weighted least squares estimation method works correctly, we estimated the parameters based on samples that were drawn from a known distribution. We drew 100 samples, each  with 100,000 data points, from an exponentiated Weibull distribution with parameters $\alpha = 1$, $\beta = 1$ and $\delta = 2$. The estimated parameters were $\hat{\alpha} = 0.996 \pm 0.067$, $\hat{\beta} = 0.998 \pm 0.033$ and $\hat{\delta} = 2.023 \pm 0.183$ ($N=100$; Fig. \ref{fig:evaluation-of-implementation}) where the values after the $\pm$-sign represent standard deviations.

\section{Comparison with gamma and beta distributions}
\label{sec:gamma-and-beta-distribution}
Two additional 3-parameter distributions were tested: The generalized gamma distribution that was proposed by Ochi \cite{Ochi1992},
\begin{equation}
    f(x) = \dfrac{c}{\Gamma(m)} \lambda^{cm} x^{cm -1} \exp[-(\lambda x)^c],
\end{equation}
and a 3-parameter beta distribution of the second kind that was proposed by Ferreira and Soares \cite{Ferreira1999},
\begin{equation}
    f(x) = \dfrac{\alpha}{B(k, n -k + 1)} \dfrac{(\alpha x)^{n - k}}{(1 + \alpha x)^{n +1}}.
\end{equation}
These distributions were fitted to the six datasets using MLE and overall mean absolute error was calculated. The errors were 0.0317$\pm$0.0203\,m and 0.0294$\pm$0.0177\,m ($N$\,=\,6) for the gamma distribution and the beta distribution, respectively (Tab. \ref{tab:mae-all-models}).
\begin{table*}[]
    \centering
    \begin{tabular}{l l l l l l l l }
         & \multicolumn{6}{l}{Dataset}\\
         Distribution & $A$ & $B$ & $C$ & $D$ & $E$ & $F$ & Mean and standard dev. \\
         \hline
         Translated Weibull  & 0.0941 & 0.0532 & 0.0492 & 0.0662 & 0.0604 & 0.0964 & 0.0699$\pm$0.0205\\
         Exponentiated Weibull & \textbf{0.0105} & \textbf{0.0219} & 0.0252 & 0.0241 & 0.0174 & 0.0561 & \textbf{0.0259$\pm$0.0158}\\
         Generalized gamma     & 0.0644 & 0.0339 & \textbf{0.0150} & \textbf{0.0205} & \textbf{0.0115} & \textbf{0.0447} & 0.0317$\pm$0.0203\\
         3-parameter beta      & 0.0112 & 0.0256 & 0.0273 & 0.0308 & 0.0190 & 0.0626 & 0.0294$\pm$0.0177 \\
    \end{tabular}
    \caption{\textbf{Overall mean absolute error in m of the four tested 3-parameter distributions.} All distributions were fitted using maximum likelihood estimation. Bold letters indicate the lowest error for the particular dataset.}
    \label{tab:mae-all-models}
\end{table*}

\bibliographystyle{elsarticle-num}

\bibliography{dissertation}

\end{document}